# On Differential Modulation in Downlink Multiuser MIMO Systems

Fahad Alsifiany*, Aissa Ikhlef†, and Jonathon Chambers*
*ComS²IP Group, School of Electrical and Electronic Engineering, Newcastle University, NE1 7RU, UK
{f.a.n.alsifiany2, jonathon.chambers}@ncl.ac.uk
†School of Engineering and Computing Sciences, Durham University, DH1 3LE, UK
aissa.ikhlef@durham.ac.uk

*Abstract*—In this paper, we consider a space time block coded multiuser multiple-input multiple-output (MU-MIMO) system with downlink transmission. Specifically, we propose to use downlink precoding combined with differential modulation (DM) to shift the complexity from the receivers to the transmitter. The block diagonalization (BD) precoding scheme is used to cancel co-channel interference (CCI) in addition to exploiting its advantage of enhancing diversity. Since the BD scheme requires channel knowledge at the transmitter, we propose to use downlink spreading along with DM, which does not require channel knowledge neither at the transmitter nor at the receivers. The orthogonal spreading (OS) scheme is employed in order to separate the data streams of different users. As a space time block code, we use the Alamouti code that can be encoded/decoded using DM thereby eliminating the need for channel knowledge at the receiver. The proposed schemes yield low complexity transceivers while providing good performance. Monte Carlo simulation results demonstrate the effectiveness of the proposed schemes.

## I. INTRODUCTION

Future wireless systems require effective transmission techniques to support high data rate and reliable communications. As such, a potential technique to utilize as part of multiple antenna systems to enhance system diversity is space-time block code (STBC) [1]. In the multiuser multiple-input multiple-output MU-MIMO downlink, transmit diversity gain can be maximized by using downlink transmission techniques such as transmit precoding, e.g., block diagonalisation (BD), and transmit spreading, such as the orthogonal spreading (OS) scheme. These techniques allow the MU-MIMO channels to be decomposed into parallel single user non-interfering channels, and hence eliminate co-channel interference (CCI) [2], [3].

For the MU-MIMO downlink, the availability of channel state information (CSI) at the transmitter makes it possible for the precoder to precancel the CCI at each user. The authors in [2] proposed a framework that uses BD to cancel the CCI and assumed full CSI knowledge at the transmitter. The CSI between the transmitter and the receivers is estimated at the receivers then fed back to the transmitter. This leads to increased complexity of the receivers. In [3], the authors proposed a method that combines the precoding technique in [2] and the Alamouti STBC. The proposed method provides a substantial gain in terms of spatial diversity with a low decoding complexity. However, for the decoding process, each receiver still needs to know the composite channel formed by the precoder and the channel in order to coherently decode the Alamouti STBC. In practice, each receiver acquires the composite channel by direct estimation.

The prior focus of STBC MU-MIMO downlink transmission techniques has been on cases where CSI is available at the receivers and transmitter. However, for some systems, due to high mobility and the cost of channel training and estimation, CSI acquisition is impossible [4]. One alternative method for such systems is differential modulation (DM). In this work, the use of DM for downlink transmission in a MU-MIMO system is considered. Specifically, we show how to use DM combined with the BD and OS schemes. Furthermore, DM is considered for both schemes based on the Alamouti STBC in order to eliminate the need for estimating the composite channels formed by the precoders and the channels at the receivers. In the BD scheme, the use of DM is to simplify the complexity of the receivers by eliminating the need for CSI as well as to cancel CCI. In particular, in order to have low complexity receivers, it is assumed that the channels are estimated at the transmitter, since it can tolerate more complexity compared to the receivers. Once the channels are estimated at the BS, the transmitter computes the precoder as in [2], [3]. However, since the BD scheme still requires CSI at the transmitter, a downlink OS scheme combined with DM is proposed. In the OS scheme, unlike the BD scheme, the transmitter does not require any knowledge of the CSI to separate the data streams of multiple users [5], [6]. Therefore, implementing the OS scheme with the DM will result in a system that does not need CSI at either ends. The proposed schemes facilitate the pre-cancelling of CCI, enhance diversity, as well as achieve a low complexity transmitter and receivers. Moreover, transmission overhead is significantly reduced using the proposed scheme, since neither feedback nor the estimation of the composite channels are required.

The rest of the paper is organized as follows. Section II introduces the system model of STBC MU-MIMO. Section III describes downlink transmission for interference cancellation. Section IV presents DM-STBC in a MU-MIMO system with downlink transmission. In Section V, the simulation results are shown. Finally, conclusions are drawn in Section VI.

## II. SYSTEM MODEL

Consider a MU-MIMO downlink broadcast channel where the base station (BS) transmits multiple streams to $K$ users (e.g., mobile stations), as shown in Fig. 1. The BS has $N_t$ transmit antennas and each user has $N_k$, $k = 1, \cdots, K$, receive antennas. The total number of receive antennas for all users is $N_r$, i.e., $N_r = \sum_{k=1}^{K} N_k$.

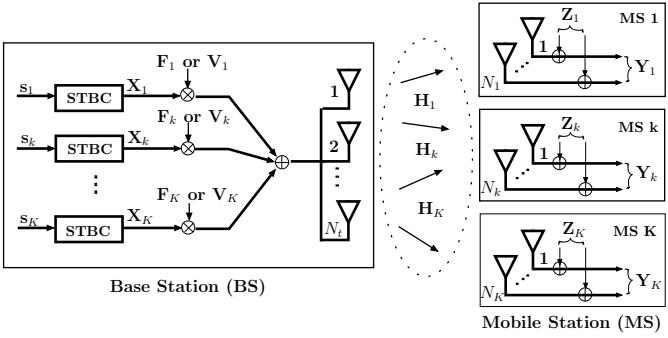

Fig. 1. STBC MU-MIMO downlink transmission system.

### A. Channel Model

The channel matrix $\mathbf{H}_k \in \mathbb{C}^{N_k \times N_t}$ for each user $k$ is a Rayleigh flat fading matrix given by

$$\mathbf{H}_k = \begin{bmatrix} h_{1,1}^{(k)} & \cdots & h_{1,N_t}^{(k)} \\ \vdots & \ddots & \vdots \\ h_{N_k,1}^{(k)} & \cdots & h_{N_k,N_t}^{(k)} \end{bmatrix} = \begin{bmatrix} \mathbf{h}_1^{(k)} \\ \vdots \\ \mathbf{h}_{N_k}^{(k)} \end{bmatrix}, \quad (1)$$

where the element $h_{i,j}^{(k)}$ is the channel coefficient between the $j$th transmit antenna and the $i$th receive antenna of user $k$, and $\mathbb{C}$ denotes the set of complex numbers. It is assumed that the channel coefficients are quasi-static over $T$ transmission slots. The elements of $\mathbf{H}_k$ are independent and identically distributed (i.i.d.) complex Gaussian random variables with zero mean and unit variance, i.e., $\mathcal{CN}(0, 1)$.

### B. Space-Time Block Coding - Alamouti Code

The multiple data streams $\mathbf{s}_k$ for each user are encoded by the Alamouti encoder to generate the STBC codeword. Let $\mathbf{X}_k \in \mathbb{C}^{2 \times 2}$, $k = 1, \cdots, K$, be the transmitted Alamouti STBC signal, satisfying the following condition [3], [7]:

$$\mathbf{X}_k^H \mathbf{X}_k = \mathbf{X}_k \mathbf{X}_k^H = \mathbf{I}_2. \quad (2)$$

The generator matrix for the Alamouti code is given as

$$\mathbf{X}_k = \frac{1}{\sqrt{2}} \begin{bmatrix} s_{1,k} & -s_{2,k}^* \\ s_{2,k} & s_{1,k}^* \end{bmatrix}, \quad (3)$$

where $s_{1,k}$ and $s_{2,k} \in \mathbb{Z}$ are the two input symbols to the Alamouti STBC encoder for user $k$. $\mathbb{Z}$ and $(.)^H$ denote the constellation set and the Hermitian operator, respectively.

## III. DOWNLINK TRANSMISSION FOR INTERFERENCE CANCELLATION

In this section, two different methods are used to cancel CCI in downlink transmission. The first scheme, referred to as the BD scheme, is suitable for the case where the CSI is available at the transmitter and the second scheme, referred to as the OS scheme, is suitable for the case where the CSI is not available at the transmitter.

### A. BD Scheme

The received signal $\mathbf{Y}_k^{(\text{BD})} \in \mathbb{C}^{N_k \times 2}$ at the $k$th user can be expressed as

$$\begin{aligned} \mathbf{Y}_k^{(\text{BD})} &= \mathbf{H}_k \mathbf{F}_k \mathbf{X}_k + \mathbf{H}_k \sum_{j=1, j \neq k}^{K} \mathbf{F}_j \mathbf{X}_j + \mathbf{Z}_k \\ &= \mathbf{H}_k \mathbf{F}_k \mathbf{X}_k + \mathbf{P}_k + \mathbf{Z}_k \end{aligned}, \quad (4)$$

where $\mathbf{F}_k \in \mathbb{C}^{N_t \times 2}$ is the precoding matrix, $\mathbf{Z}_k \in \mathbb{C}^{N_k \times 2}$ is an AWGN noise matrix. $\mathbf{P}_k \in \mathbb{C}^{N_k \times 2}$ is the CCI component at the $k$th user. Note that, at the BS, the precoding matrix $\mathbf{F}_k$ for the $k$th user is multiplied by the symbol vector and added to the precoded signals from the other users to produce the composite transmitted matrix, i.e., $\sum_{k=1}^{K} \mathbf{F}_k \mathbf{X}_k$.

The BD method employs precoding matrices $\mathbf{F}_k$, $k = 1, \cdots, K$, to completely suppress the CCI at the receivers. To cancel the CCI, the following constraint should be satisfied [2], [3]

$$\mathbf{H}_j \mathbf{F}_k = 0, \quad j, k = 1, ..., K, \ j \neq k. \quad (5)$$

Let $\bar{\mathbf{H}}_k \in \mathbb{C}^{\bar{N}_k \times N_t}$, where $\bar{N}_k = N_r - N_k$, denote the channel matrix for all $K$ users excluding the $k$th user's channel, which is defined as

$$\bar{\mathbf{H}}_k = \begin{bmatrix} \mathbf{H}_1^H & \cdots & \mathbf{H}_{k-1}^H & \mathbf{H}_{k+1}^H & \cdots & \mathbf{H}_K^H \end{bmatrix}^H. \quad (6)$$

Therefore, the zero-interference constraint in (5) is re-expressed as

$$\bar{\mathbf{H}}_k \mathbf{F}_k = 0, \quad k = 1, ..., K. \quad (7)$$

According to [3], to satisfy (7), one solution is to construct $\mathbf{F}_k$ as

$$\mathbf{F}_k = (\mathbf{I} - \bar{\mathbf{H}}_k^\dagger \bar{\mathbf{H}}_k) \mathbf{\Phi}_k, \quad (8)$$

where $\mathbf{\Phi}_k \in \mathbb{C}^{N_t \times 2}$ is an eigenmode selection matrix, and $(.)^\dagger$ denotes the pseudo-inverse. The magnitude, i.e, the vector norm of the precoding matrix $\mathbf{F}_k$ has to be unity to ensure a constant transmission power for the $k$th user, i.e.,

$$\mathbf{F}_k^H \mathbf{F}_k = \mathbf{I}_2, \quad k = 1, \cdots, K. \quad (9)$$

Therefore, to satisfy (9), the unitary $\mathbf{F}_k$ matrix can be constructed as a linear combination of the column space spanning vectors of $(\mathbf{I} - \bar{\mathbf{H}}_k^\dagger \bar{\mathbf{H}}_k)$, which can be obtained by the Gram-Schmidt orthogonalization (GSO), or the standard QR decomposition. In this paper, QR decomposition is used for its simplicity.

To compute $\mathbf{\Phi}_k$, a singular value decomposition (SVD) of $\mathbf{H}_k (\mathbf{I} - \bar{\mathbf{H}}_k^\dagger \bar{\mathbf{H}}_k)$ is performed. This is done by selecting the two singular vectors corresponding to the two largest singular values of $\mathbf{H}_k(\mathbf{I} - \bar{\mathbf{H}}_k^\dagger \bar{\mathbf{H}}_k)$. The resulting received signal for the $k$th user after cancelling out the CCI is given by

$$\mathbf{Y}_k^{(\text{BD})} = \mathbf{H}_k \mathbf{F}_k \mathbf{X}_k + \mathbf{Z}_k = \breve{\mathbf{H}}_k \mathbf{X}_k + \mathbf{Z}_k, \quad (10)$$

where $\breve{\mathbf{H}}_k \in \mathbb{C}^{N_k \times 2}$ is the effective channel for user $k$.

### B. OS Scheme

In the OS case, the received signal matrix $\mathbf{Y}_k^{(\text{OS})} \in \mathbb{C}^{N_k \times KN_t}$ for the $k$th user is given by [5]

$$\mathbf{Y}_k^{(\text{OS})} = \mathbf{H}_k \mathbf{X}_k \mathbf{V}_k + \mathbf{H}_k \sum_{j=1, j \neq k}^{K} \mathbf{X}_j \mathbf{V}_j + \mathbf{Z}_k, \quad (11)$$

where $\mathbf{V}_k \in \mathbb{C}^{N_t \times KN_t}$ is the orthogonal spreading matrix for user $k$, $\mathbf{Z}_k \in \mathbb{C}^{N_k \times KN_t}$ is an AWGN noise matrix. The composite transmitted matrix is $\sum_{k=1}^{K} \mathbf{X}_k \mathbf{V}_k$. Note that, in order to apply Alamouti STBC along with the orthogonal spreading code, the number of transmit antennas at the BS has to be limited to two, i.e, $N_t = T = 2$.

In the OS scheme, each user is assigned a unique orthogonal spreading code to separate the data of the users at the receivers. The STBC codeword for each user is multiplexed by its own specific spreading code and then transmitted. As in the BD method case, to eliminate CCI, the spreading code matrix has to obey the following conditions

$$\mathbf{V}_k \mathbf{V}_k^H = \mathbf{I}_{N_t}, \quad k = 1, ..., K. \quad (12)$$
$$\mathbf{V}_j \mathbf{V}_k^H = 0, \quad k, j = 1, ..., K, \text{ and } j \neq k. \quad (13)$$

The OS code for each user can be constructed as a submatrix of the Hadamard matrix, or from a discrete Fourier transform (DFT) matrix. Hadamard matrices are of interest because of their simplicity. Hadamard codes are a set of orthogonal codes which are built repeatedly from the basic building block

$$\mathbf{A}_2 = \frac{1}{\sqrt{2}} \begin{bmatrix} +1 & +1 \\ +1 & -1 \end{bmatrix} \quad (14)$$

according to

$$\mathbf{A}_{2^{n+1}} = \frac{1}{\sqrt{2^{n+1}}} \begin{bmatrix} \mathbf{A}_{2^n} & \mathbf{A}_{2^n} \\ \mathbf{A}_{2^n} & -\mathbf{A}_{2^n} \end{bmatrix}, \quad (15)$$

where the dimension of the Hadamard matrix in (15) is $2^{n+1} \times 2^{n+1}$. Note that in our case $2^{n+1} = KN_t$.

Due to the orthogonality of the spreading matrices used at the transmitter, at each receiver, the original information signal is retrieved by despreading the received signal with the synchronized duplicate of the spreading code. Therefore, the received signal matrix $\mathbf{Y}_k^{(\text{OS})}$ in (11) for the $k$th user is despread by multiplying it with $\mathbf{V}_k^H$, which yields

$$\hat{\mathbf{Y}}_k^{(\text{OS})} = \mathbf{Y}_k^{(\text{OS})} \mathbf{V}_k^H = \mathbf{H}_k \mathbf{X}_k + \hat{\mathbf{Z}}_k, \quad (16)$$

where $\hat{\mathbf{Y}}_k^{(\text{OS})} \in \mathbb{C}^{N_k \times N_t}$ is the despread received signal, and $\hat{\mathbf{Z}}_k \in \mathbb{C}^{N_k \times N_t}$ is the despread AWGN noise.

### C. Complexity Analysis

In this section, the computational complexity with the notion of flops is introduced here, where flops denotes the floating point operation. At the transmitter, the BD scheme uses the spatial dimension to cancel CCI, whereas the OS scheme uses the time dimension. In the BD scheme, in order to cancel CCI completely, the system must satisfy [2], [3]

$$N_t \geq \left( \sum_{j=1, j \neq k}^{K} N_j + 2 \right). \quad (17)$$

The complexity of the BD scheme is mainly based on pseudo-inverse $\bar{\mathbf{H}}_k^\dagger = \bar{\mathbf{H}}_k^H \left( \bar{\mathbf{H}}_k \bar{\mathbf{H}}_k^H \right)^{-1}$, and the QR decomposition of $(\mathbf{I} - \bar{\mathbf{H}}_k^\dagger \bar{\mathbf{H}}_k)$. The complexity of both the pseudo-inverse operation and the QR decomposition follows [8], [9]

$$\mathcal{O}\left( KN_t \left( \sum_{j=1, j \neq k}^{K} N_j \right)^2 \right). \quad (18)$$

In the OS scheme, the precoder is independent from the number of receive antennas. Thus, the complexity of the OS scheme is only based on Hadamard matrix construction which is already given. Hence, it does not incur any computational complexity. Obviously, the OS scheme has lower computational complexity than the BD scheme, but in terms of throughput, the OS scheme throughput is $K$ times smaller than that of the BD scheme. Note that, the computational complexity at the receiver side for both schemes is the same, and we will explore more about the DM decoder in the following section.

## IV. DIFFERENTIAL STBC FOR MU-MIMO WITH DOWNLINK TRANSMISSION

In this section, the differential encoding and decoding process for downlink transmission in a MU-MIMO system is discussed. In particular, this section demonstrates how to use the BD and OS schemes in differential STBC MU-MIMO systems.

### A. Differential Encoding

The particular encoding algorithm utilized for DM builds upon the works in [7], [10]. The algorithm requires that unitary STBCs such as the Alamouti code are used. In the encoding process, the $\mathbf{X}_0$ matrix is used as a reference code, in which the transmitted matrix for the initial block of each user $k$ is set to be identity as

$$\mathbf{X}_{0,k} = \mathbf{I}_T, \quad k = 1, \cdots, K. \quad (19)$$

Then, for the BD scheme, the unitary Alamouti STBC matrices are encoded differentially for the subsequent blocks as follows

$$\mathbf{B}_n^{(\text{BD})} = \sum_{k=1}^{K} \mathbf{F}_k \left( \prod_{i=0}^{n} \mathbf{X}_{i,k} \right), \quad n = 0, ..., N. \quad (20)$$

For the OS scheme, the encoding process is as follows

$$\mathbf{B}_n^{(\text{OS})} = \sum_{k=1}^{K} \left( \prod_{i=0}^{n} \mathbf{X}_{i,k} \mathbf{V}_k \right), \qquad n = 0, ..., N, \quad (21)$$

where $\mathbf{B}_n^{(q)}$, $q \in \{\text{BD, OS}\}$, is the $n$th encoded block, $N+1$ is the total number of encoded signal blocks, and $\mathbf{F}_k$ and $\mathbf{V}_k$ represent the precoding matrix and spreading matrix for user $k$, respectively.

The performance of the differential modulation system depends on the length of time over which the channel coefficients remain constant. Ordinarily, the reference (known) symbol $\mathbf{X}_{0,k}$ must be sent periodically, based on the channel coherence time. Accordingly, generating the downlink precoding matrix $\mathbf{F}_k$ or the downlink spreading matrix $\mathbf{V}_k$ for the new channel coefficient matrix only needs to be done when there are new channel coefficients.

*B. Differential Decoding*

For the MU-MIMO downlink system, the differential transmissions are implemented in blocks, in which each user $k$ receives the sum of all the transmit waveforms of other users; then the received signal blocks for each user must be detected independently. Thus, if $\mathbf{G}_k$ denotes the matrix having all $N+1$ received signal blocks for the $k$th user, i.e.,

$$\mathbf{G}_k = [\mathbf{Y}_{0,k} \ \mathbf{Y}_{1,k} \ \cdots \ \mathbf{Y}_{N,k}], \quad (22)$$

then the received signal block at the $k$th user during the $n$th iteration block, i.e., $\mathbf{Y}_{n,k}$ can be expressed as

$$\mathbf{Y}_{n,k} = \mathbf{H}_k \mathbf{B}_n^{(q)} + \mathbf{Z}_{n,k}, \qquad n = 0, ..., N, \quad (23)$$

where $q \in \{\text{BD, OS}\}$, and $\mathbf{Z}_{n,k}$ is the $k$th user AWGN noise during the $n$th block. For DM encoding, it is assumed that the channel matrix $\mathbf{H}_k$ changes slowly (channel coherence time is large enough) and extends over several matrix transmission periods. In such a case, the BS transmission starts with a reference matrix, followed by several information matrices. When encoding using (20) or (21), the decoding process for $\mathbf{X}_{n,k}$ would be according to the last two blocks of $\mathbf{G}_k$ as in the following notation [7], [10]

$$\mathbf{G}_k = \left[ \underbrace{\mathbf{Y}_{0,k} \mathbf{Y}_{1,k}}_{} \cdots \underbrace{\mathbf{Y}_{n-1,k} \mathbf{Y}_{n,k}}_{} \cdots \underbrace{\mathbf{Y}_{N-1,k} \mathbf{Y}_{N,k}}_{} \right]. \quad (24)$$

For the BD method, to make this more explicit, define

$$\mathbf{Y}_{n,k} \triangleq \begin{bmatrix} \mathbf{Y}_{n-1,k} \\ \mathbf{Y}_{n,k} \end{bmatrix} \triangleq \begin{bmatrix} \mathbf{H}_k \mathbf{B}_{n-1}^{(q)} + \mathbf{Z}_{n-1,k} \\ \mathbf{H}_k \mathbf{B}_n^{(q)} + \mathbf{Z}_{n,k} \end{bmatrix}, \quad (25)$$

and recall from (5) that the interference of other users is suppressed, thus the two blocks in (25) become a single user block matrix as

$$\mathbf{Y}_{n,k} \triangleq \begin{bmatrix} \mathbf{H}_k \mathbf{F}_k \mathbf{X}_{n-1,k} + \mathbf{Z}_{n-1,k} \\ \mathbf{H}_k \mathbf{F}_k \mathbf{X}_{n-1,k} \mathbf{X}_{n,k} + \mathbf{Z}_{n,k} \end{bmatrix}. \quad (26)$$

The code matrices that affect $\mathbf{Y}_{n,k}$ are

$$\mathbf{D}_{X_{n,k}} = \begin{bmatrix} \mathbf{X}_{n-1,k} \\ \mathbf{X}_{n-1,k} \mathbf{X}_{n,k} \end{bmatrix}. \quad (27)$$

Assuming that $N_t = T$, and using these results, as well as (2) and (9), the matrices in (27) can be expressed as

$$\mathbf{D}_{X_{n,k}}^H \mathbf{D}_{X_{n,k}} = 2\mathbf{I}_{N_t}, \quad (28)$$

therefore, these matrices represent unitary block codes. When $\mathbf{X}_{n-1,k}$ is known to the receiver, the optimal decoder for this block is the quadratic receiver as [10]

$$\hat{\mathbf{X}}_{n,k} = \arg \max_{\mathbf{X}_{n,k}} \text{trace} \left\{ \mathbf{Y}_{n,k} \mathbf{D}_{X_{n,k}} \mathbf{D}_{X_{n,k}}^H \mathbf{Y}_{n,k}^H \right\}. \quad (29)$$

Since we have

$$\mathbf{D}_{X_{n,k}} \mathbf{D}_{X_{n,k}}^H = \begin{bmatrix} \mathbf{I}_T & \mathbf{X}_{n,k}^H \\ \mathbf{X}_{n,k} & \mathbf{I}_T \end{bmatrix}, \quad (30)$$

the decoder in (29) can be re-written as follows [10], [7]

$$\hat{\mathbf{X}}_{n,k} = \arg \max_{\mathbf{X}_{n,k}} \text{trace} \left\{ \begin{bmatrix} \mathbf{Y}_{n-1,k} \\ \mathbf{Y}_{n,k} \end{bmatrix} \begin{bmatrix} \mathbf{I}_T & \mathbf{X}_{n,k}^H \\ \mathbf{X}_{n,k} & \mathbf{I}_T \end{bmatrix} \begin{bmatrix} \mathbf{Y}_{n-1,k} \\ \mathbf{Y}_{n,k} \end{bmatrix}^H \right\}$$

$$= \arg \max_{\mathbf{X}_{n,k}} \Re \left\{ \text{trace} \left\{ \mathbf{X}_{n,k} \mathbf{Y}_{n,k}^H \mathbf{Y}_{(n-1),k} \right\} \right\}, \quad (31)$$

where $\Re(.)$ denotes the real part, and $\text{trace}(.)$ denotes the trace of a matrix. Similarly, the equivalent differential decoder for the OS scheme can be constructed. Note that when the CSI is available at the receiver, the standard Alamouti decoder is used before the maximum likelihood (ML) detection is implemented upon the combined signals.

## V. SIMULATIONS RESULTS AND DISCUSSION

In this section, the performance of the differential and coherent Alamouti STBC for MU-MIMO downlink transmission is examined. Alamouti codes with QPSK are used throughout the simulation.

Fig. 2 plots the symbol error rate (SER) for coherent modulation (CM) and DM with one receive antenna per user. For BD scheme, the performance curve is plotted for a single user system with 2 transmit antennas at the BS and a four-user system with 5 transmit antennas at the BS. For OS scheme, the number of transmission antenna has been set to be always two against 1 and 4 users. We observe that CM and DM for both BD and OS schemes achieve the same performance as a single-user STBC-MISO link; that is, CCI is completely eliminated and full diversity is achieved with the Alamouti code. Ordinarily, the differential detection underperforms the coherent detection by about 3 dB.

Fig. 3 illustrates the results of repeating the experiment with two receive antennas per user. Similarly, the MU-MIMO system of CM and DM for both schemes behave as a single user STBC-MIMO link, but with better performance than the one receive antenna per user system. For BD scheme, CCI elimination requires that the number of transmit antennas is sufficient to achieve full diversity with the given number of receive antennas, so $N_t = 8$ is chosen. For OS scheme, we have got the same performance but with fixed number of transmit antennas, e.g., $N_t = 2$. Consequently, unlike BD scheme, the number of receive antenna per user is independent from the number of transmission antenna.

Fig. 4 shows the performance of exploiting DM combined with BD and OS schemes with three receive antennas per user. The high mobility and multipath propagation may result in multiple access interference (MAI) in OS scheme and imperfect channel estimation in BD scheme, which destroy the orthogonality of the precoders. Hence, Fig. 4 also shows the impact of possible errors in both schemes. For OS scheme, the error spreading matrix for user 1 is $\bar{\mathbf{V}}_1 = \mathbf{V}_1 + \alpha \mathbf{V}_2$, where $\alpha$ is the error coefficient [5]. The values of $\alpha$ are chosen to be $0.1, 0.2$, respectively. For BD scheme, imperfect channel matrix at the BS for user 1 is $\ddot{\mathbf{H}}_1 = \mathbf{H}_1 + \mathbf{E}_1$, where $\mathbf{H}_1$ is the perfect channel estimate for user 1 and $\mathbf{E}_1$ is the error matrix [3]. Entries of $\mathbf{E}_1$ are i.i.d. Gaussian variables with distribution zero mean and covariance of $\sigma^2$. The values of $\sigma$ are chosen to be $0.1$ and $0.2$. From Fig. 4, it is clear that the OS is more robust against errors compared to the BD scheme.

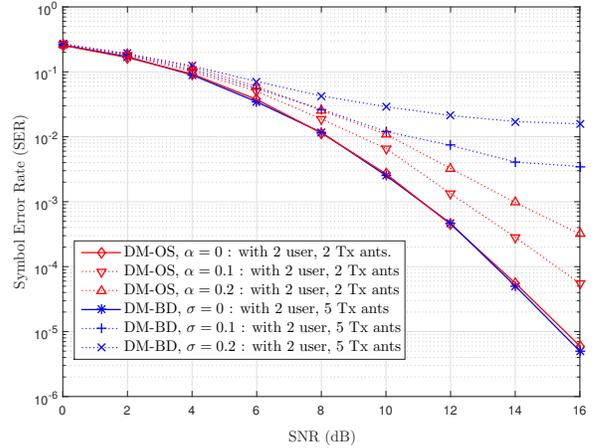

Fig. 4. SER performance of differential detection system using BD and OS schemes for $N_k = 3$ with the impact of precoding errors on user 1.

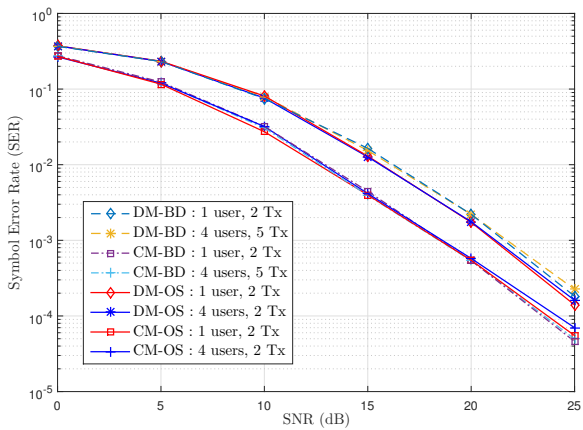

Fig. 2. SER performance of MU-MIMO STBC downlink precoding with coherent and differential detection using BD and OS schemes for $N_k = 1$.

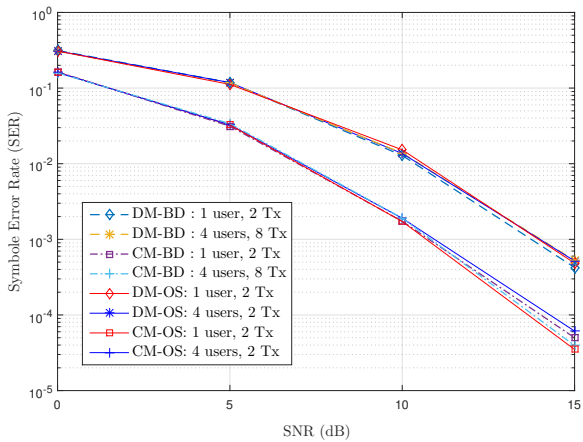

Fig. 3. SER performance of MU-MIMO STBC downlink precoding with coherent and differential detection using BD and OS schemes for $N_k = 2$.

## VI. CONCLUSION

In this paper, a low complexity differential STBC scheme for MU-MIMO with downlink transmission has been proposed. In particular, DM combined with either the BD scheme or the OS scheme overcame the need for CSI at the receivers as well as cancelled CCI. On the other hand the use of STBC can achieve full diversity without needing CSI at the transmitter. It has been demonstrated that implementing the BD scheme with DM will establish a system that does not need CSI at the receivers to decode the signals, while combining the OS scheme with DM will establish a system that requires CSI at neither the transmitter nor at the receivers. The differential modulation for both systems loses typically 3dB in performance relative to the coherent detection method, but this is offset by the reduction in complexity of the receivers and the transmitter. The BD scheme is more complex than the OS scheme; however, the BD scheme has a higher throughput. Moreover, it was shown that the OS is more robust against precoding errors compared to the BD scheme.